\documentclass[12pt,preprint]{aastex}

\lefthead{Dallacasa et al.}
\righthead{Deep 1.4 GHz follow up of Abell 521}

\begin{document}


\title{DEEP 1.4 GHZ FOLLOW UP OF THE STEEP SPECTRUM RADIO HALO IN ABELL 521}

\author{D.Dallacasa\altaffilmark{1}, G.Brunetti\altaffilmark{2}, 
S.Giacintucci\altaffilmark{3}, R.Cassano\altaffilmark{2},
T.Venturi\altaffilmark{2}, G.Macario\altaffilmark{1,2}, 
N.E.Kassim\altaffilmark{4}, W.Lane\altaffilmark{4}, G.Setti\altaffilmark{1,2}}

\altaffiltext{1}{Dipartimento di Astronomia, Universit\'a di
Bologna, via Ranzani 1, I--40127 Bologna, Italy}
\altaffiltext{2}{INAF- Istituto di Radioastronomia, via Gobetti 101,
I--40129 Bologna, Italy}
\altaffiltext{3}{Harvard-Smithsonian Center for Astrophysics,
  Cambridge, Massachusetts 02138, USA} 
\altaffiltext{4}{Naval Research Laboratory, Code 7213, Washington DC
20375-5320, USA}

\begin{abstract}
In a recent paper we reported on the discovery of a radio halo with
very steep spectrum in the merging galaxy cluster Abell 521 through
observations with the {\it Giant Metrewave Radio Telescope} (GMRT). 
We showed that the steep spectrum of the halo is inconsistent with
a secondary origin of the relativistic electrons and supports a 
turbulent acceleration scenario.
At that time, due to the steep spectrum, the available observations 
at 1.4 GHz (archival NRAO-{\it Very Large Array $-$ VLA $-$ }
CnB-configuration data) were not adequate to 
accurately determine the flux density associated with the radio halo. 
In this paper we report the detection at 1.4 GHz of the radio halo in 
Abell 521 using deep VLA observations in
the D-configuration.  We use these new data to confirm the steep-spectrum
of the object.  
We consider Abell 521 the prototype of a population of very-steep
spectrum halos.
This population is predicted assuming that turbulence plays an important 
role in the acceleration of relativistic particles in galaxy clusters, 
and we expect it will be unveiled by future surveys at low frequencies 
with the LOFAR and LWA radio telescopes.
\end{abstract}


\section{INTRODUCTION}

Galaxy clusters are the largest virialized structures in the Universe.
They contain about $10^{15}$ M$_{\odot}$ in the form of hot gas,
galaxies, dark matter and non-thermal components.
The latter, such as magnetic fields and high energy particles, may
play key roles by controlling transport processes in the
intergalactic medium (IGM) and are sources of additional pressure and
energy support (Ryu et al. 2003; Schekochihin et al. 2005; 
Lazarian 2006). 

The origin of the non-thermal particles is likely connected with the
cluster formation process. A fraction of the energy is dissipated in
the form of shocks and turbulence during cluster mergers. The
accretion of matter may be channelled into the acceleration of
particles, thus leading to a complex population of non-thermal primary 
particles in the IGM producing synchrotron and inverse Compton
(IC) radiation (Roettiger et al. 1999; Sarazin 1999; Ryu et al. 2003;
Brunetti et al. 2004; Brunetti \& Lazarian 2007;  Hoeft \& Bruggen
2007; Pfrommer et al. 2008). 
Relativistic protons are expected to be the dominant non-thermal
particle component (e.g. Blasi et al. 2007); collisions between
relativistic and thermal protons in the IGM inject secondary electrons
and neutral pions that can also produce synchrotron (plus inverse
Compton) and gamma-ray emission respectively (Voelk et al. 1996;
Berezinsky et al. 1997; Blasi \& Colafrancesco 1999; Pfrommer \&
Ensslin 2004; Brunetti \& Blasi 2005). 

The most prominent examples of non-thermal activity in galaxy clusters
are giant radio halos.
They are low surface brightness, Mpc-scale diffuse sources of
synchrotron radiation produced by relativistic electrons moving into
$\mu$G magnetic fields, which are frozen in the hot cluster gas
(e.g. Feretti~2003).  
Two main mechanisms may account for the relativistic electrons
emitting the observed radiation: either injection of secondary
electrons by collisions between relativistic and thermal protons in
the IGM (Dennison 1980; Blasi \& Colafrancesco 1999), or in situ
acceleration of relativistic electrons by shocks and turbulence
generated during cluster mergers (Brunetti et al. 2001; Petrosian
2001). 
Recent observations of radio halos suggest that MHD turbulence,
injected during cluster mergers, may play an important role in the
process of particle acceleration (e.g., reviews by Brunetti 2008;
Ferrari et al. 2008; Cassano 2009). 
Spectral studies are extremely important to address this issue: the
detection of radio halos with very steep spectrum, $\alpha > 1.5$ (S
$\sim \nu^{-\alpha}$), 
supports stochastic particle acceleration (e.g. due to MHD turbulence)
and disfavours a secondary origin of the electrons, due to the large 
proton-energy budget that this scenario requires (Brunetti 2004).

In a recent paper we reported on the discovery of a radio halo with
very steep synchrotron spectrum, $\alpha \approx 2$ , associated with
the merging galaxy cluster Abell 521 (Brunetti et al. 2008).  
In this paper we report on 1.4 GHz follow up deep observation of the
radio halo in Abell 521 with the VLA in the D-configuration. We
successfully detected  the radio halo and derived the overall  
spectrum between 240 MHz and 1.4 GHz.
In Sect.~2 we report on these new observations, in Sect.~3 we discuss
the results, while in Sect.~4 we give our conclusions.

A $\Lambda$CDM concordance model with $H_0=70$ km s$^{-1}$ Mpc$^{-1}$,
$\Omega_m=0.3$ and $\Omega_{\Lambda}=0.7$ is used.

\section{ABELL 521}

Abell 521 is an X-ray luminous ($8.2 \times 10^{44}$ erg s$^{-1}$ in
the band 0.1--2.4 keV) and massive ($\approx 2 \times 10^{15}$
M$_{\odot}$) galaxy cluster at z=0.247 with ongoing multiple merging
episodes (Arnaud et al. 2000; Ferrari et al. 2003). 
A radio relic is located on the south-eastern boundary of the cluster
(Ferrari et al. 2006; Giacintucci et al. 2006) and is found to
coincide with a possible shock front, generated by the recent infall
of a subcluster along the North-West/South-East direction, where
relativistic  electrons are currently accelerated (Giacintucci et
al. 2008). 

Brunetti et al. (2008) reported on the detection of a Mpc-scale radio
halo with the GMRT at 240, 330 and 610 MHz. 
The halo has an angular size of about 4$^\prime$ with a rather smooth
brightness distribution. 
Available archival VLA observations at 1.4 GHz in the BnC
configuration (project AF390), reanalized in Brunetti et al. (2008), 
lacked both the sensitivity and the short baseline coverage
to reliably detect the halo and prevented us from a proper flux density
measurement. Therefore only an upper limit on the total flux density
could be placed.
The observed integrated radio spectrum has been found very steep,
$\alpha=2$ (Brunetti et al. 2008). 
We observed A521 at 1.4 GHz with the VLA in the D-configuration whose
short spacings are well suited to study the halo and match those 
provided by the GMRT at 240 MHz, allowing a reliable measurement of
the spectral index.

\section{Observations at 1.4 GHz} 

We observed A521 on August 24 2008, with the VLA in the D configuration
for a total time on source of about 5.5 hr.
Observations were made using a multi-channel continuum mode to permit
RFI excision and prevent bandwidth smearing of field sources.  Seven
6.25 MHz channels were combined to create an effective bandiwdth of
43.75 MHz centered at 1364.9 MHz. 
The data were calibrated in a standard way, with amplitude, phase and
bandpass calibration carried out after accurate editing of raw data on
both the primary (3C48) and secondary (0423-013) calibration sources.
The accuracy of the flux density scale is within 3\% as estimated from
the variation of the gain solutions over the whole observation.  The
last hour of the observation was affected by some RFI 
covering the whole observing bandwidth; this clearly showed up when
circular polarization (Stokes V) was considered. The telescopes were
pointing at low elevation and it is likely that some signal of human
origin entered their beams. Therefore we decided to drop that part of
the observation in the imaging process.
After self-calibration, the edited D-configuration data were imaged.
The final D-configuration image has a resolution of
58$^{\prime\prime}\times$35$^{\prime\prime}$. We measured emission in
the 1 Mpc halo region as in
Brunetti et al. (2008) and found $S_{(1.4 {\rm GHz})} = 10.1$ mJy,
including the contribution from discrete sources.
In order to obtain a more sensitive image we combined our D-configuration 
data set with the archival VLA BnC-configuration data (total time
$\sim$5 hours), adopting the following procedure.
We first made an image of the A521 D-configuration region setting a
number of additional facets on all the confusing sources within the
primary beam but far from the field centre. Then we removed the clean
components of such sources from the visibilities, and averaged the
seven 6.25 MHz channels. 
The same kind of subtraction was applied to the archival data 
before proceeding with the combination of the two datasets.

A proper sampling of the short spacings is a key issue in the correct
determination of the flux density of extended and low surface
brightness sources like the halo in A521. A comparison between the
inner portion of the uv-coverage at the various frequencies is
reported in Figure 1. Visibilities corresponding to baselines shorter
than 1 k$\lambda$ have been plotted. The angular size of the halo
(around 4$^\prime$) is sampled by the visibilities shorter than about
0.9 k$\lambda$. It is clear that the data at 240 and 1400 MHz are those
better suited to determine the flux density of the halo since they
have comparable uv-coverages. At 330 MHz the density of the samples
starts to get sparse. At 610 MHz, the shortest spacings are poorly
sampled, leading to an underestimate of the total flux density of the
halo.

The final images from the combined dataset were obtained after a
number of phase-only self-calibrations. We first made an image using
baselines longer than 0.8 k$\lambda$ to find the point sources embedded
in the halo region. After choosing an appropriate weighting we
obtained an image with a resolution of
13.7$^{\prime\prime}\times$6.6$^{\prime\prime}$ (with PA = 80 $\deg$)
and 14$\mu$Jy r.m.s. noise. Eleven point sources in the central region
of A521, with peak flux density exceeding $6\sigma$, were fitted with
individual Gaussians and subtracted from the data.
A lower resolution image was then made to highlight the halo
structure and provide a good comparison to lower frequency images.
Figure 2 (right panel) shows the
30$^{\prime\prime}\times$30$^{\prime\prime}$ resolution image of the
halo at 1.4 GHz:  
the radio halo is well detected with a morphology similar to that
found at lower frequencies (240 MHz, Figure 2, left panel). The halo
becomes progressively less dominant at higher frequencies implying
that its spectrum is considerably steeper than that of the relic,
$\alpha \approx 1.5$ (Giacintucci et al. 2008). 
The halo emission has low surface brightness at 1.4 GHz: we find
$6.4\pm 0.6$ mJy of diffuse emission in the 1 Mpc-diameter circular
region (as in Brunetti et al. 2008) centered on the central cluster 
galaxy. The error considered here includes the uncertainty in the flux
density arising from the subtraction procedure of the discrete
sources. The flux density of the halo emission is about 30 percent
larger than the upper limit in Brunetti et al. (2008) that was
estimated through the injection of fake radio halos in the uv-data of
the archive BnC-VLA observations available at that time. Most likley
differences arise from the real profile of the halo, which has some
substructures and is slightly flatter than that adopted in the
modelling.   
Some weak diffuse radio emission extends slightly beyond the
region (1 Mpc in diameter), which has been used in the flux density
measurements at the various frequencies (see Brunetti et
al. 2008). Such emission is visible at all frequencies and would
increase all the flux densities of a few percent, thus not affecting
the total spectral index.

We confirm that the spectrum of the halo is very steep: we find
$\alpha = 1.86 \pm 0.08$ between 330-1400 MHz. For comparison, the
typical slope of giant radio halos is $\alpha \approx 1.2-1.3$
(e.g. Feretti et al. 2004). 
The spectrum of the halo in A521 could be even steeper if we consider 
the possible contribution to the flux density of the diffuse emission 
arising from cluster sources that are not detected in our radio maps
(and whose spectral index are generally around 0.7-0.8); this 
should be more relevant at 1.4 GHz, due to the steep spectrum of the 
radio halo. We do not expect this to be an large effect because,
based on the {\it average} radio luminosity functions of radio sources 
in X-ray luminous clusters (e.g. Branchesi et al. 2006), we estimate this 
contribution within $\approx$0.2-0.4 mJy at 1.4 GHz.

\section{DISCUSSION}

The radio spectrum of the halo is shown in Figure 3.
The measurement at 610 MHz is a factor $\approx 1.6$ below the overall
spectral shape marked by the data points at 240, 330 and 1400 MHz.
This can be explained if we consider the sparse short baselines uv-coverage
associated with the $\approx 2$ hrs observations of
the 610 MHz {\it GMRT Radio Halo Survey} (Venturi et
al. 2007,08). Such observations were aimed to reveal radio halos in
X-ray luminous clusters at intermediate redshift, and do not have 
the sensitivity and short baseline coverage to provide detailed
images of the faintest halos. 

The deep observations at 240, 330 and 1400 MHz and their excellent
uv-coverage matched to the spatial scales of the halo show a
gradual steepening of the halo spectrum with increasing frequency
(Figure 3), although the steepening cannot be large since our
observations cover only a factor of $\approx 5$ in frequency. 
Spectral steepening is a key feature of turbulent acceleration in 
merging cluster (Schlickeiser et al. 1987; Brunetti et
al. 2004). To highlight this point, Figure 3 includes an example of a
theoretical spectrum of synchrotron emission from Abell 521 in the
case of magnetosonic re-acceleration of low-energy seed relativistic
electrons (Brunetti \& Lazarian 2007). 
Besides the spectral curvature, we stress that the observation of a
very steep spectrum in these environments provides evidence for
turbulent acceleration in the IGM (e.g. Brunetti 2004, 2008; Cassano
2009). 

On the other hand, due to straightforward energy arguments, such a
very steep spectrum of the radio halo in Abell 521 is not consistent
with a secondary origin of the emitting electrons.
In the secondary electron model, the primary relativistic protons
collide with the thermal protons: they inject the secondary electrons
whose energy spectrum is $\delta = 2 \alpha$, or even slightly steeper
when the logarithmic behaviour of the p-p cross section with the
proton-energy is taken into account (e.g. Dermer 1986; Brunetti \&
Blasi 2005). 
We adopt the formalism in Brunetti \& Blasi (2005) to calculate the 
injection rate of secondary electrons assuming the same parameters for
the thermal IGM as in Brunetti et al. (2008). To match the flux
density of the radio halo measured at 330 MHz, we would need a
relativistic proton population whose energy density 
exceeds that of the thermal IGM in Abell 521, as it is derived from
observations under the assumption of a reasonable value of the
magnetic field averaged in the halo volume: $B \leq 7 \mu$G). As
remarked in Brunetti 
et al. (2008), this yields only a lower limit to the energy density of
high energy protons since, for $\delta > 3$, an additional (dominant)
contribution to the energy comes from supra-thermal particles with
kinetic energies $\leq$ 1 GeV.  

The turbulent acceleration scenario predicts that radio halos will be
more easily revealed at low frequencies (few hundred MHz), since the
electrons emitting at these frequencies can be accelerated also during
less energetic mergers, which are more frequent than major mergers in
the Universe (Cassano et al. 2006). 
These low frequency halos have a steep spectrum and they may have been
missed by present high frequency (at 1.4 GHz) surveys.  Abell 521
illustrates this point; it was not detected by earlier observations at
1.4 GHz (Ferrari et al. 2006). 
Based on evidence from low frequency GMRT data, some residual radio
emission was detected in the halo region at 1.4 GHz from a re-analysis
of those BnC-configuration data (Brunetti et al. 2008), yet it was 
also clear that a proper deep VLA observation was necessary to 
accurately image the diffuse emission associated with this cluster. 
Future surveys at low frequencies with LOFAR and LWA will enter an
unexplored territory providing a chance to detect many of these radio 
halos and unveil their properties. In this respect we believe that the
case Abell 521 provides a glimpse of what these surveys might find.

\section{CONCLUSIONS}

We report on VLA--D configuration 1.4 GHz observations of the steep
spectrum giant radio halo in the merging cluster A\,521, recently
discovered thanks to GMRT low radio frequency observations.
The combination of our new deep observations obtained with the VLA in
the D configuration with archival 1.4 GHz data (VLA, BnC
configuration), allowed us to greatly improve the uv-coverage of the
interferometric observation and to successfully detect the faint
diffuse emission associated with A\,521 also at this frequency.
The 1.4 GHz flux density of the giant radio halo, derived after
subtraction of the embedded discrete sources, confirms that the
integrated spectrum of the radio halo has a very steep slope, with
$\alpha=1.86\pm0.08$.
The very steep spectrum and the hint of spectral curvature confirm  
the idea that turbulence plays an important role in the 
acceleration of relativistic particles in this cluster. 

Our results are inconsistent with a secondary acceleration
scenario. The new observations detect $\sim$30\% more flux at 1.4 GHz
than previously constrained, implying that the spectrum of relativistic
protons in the secondary electron scenario is slightly flatter
than adopted in Brunetti et al. (2008) ($\Delta\delta \sim$ 0.3).
However, this reduces the energy budget required to match
the radio emission only by a factor of 2 - 2.5, and thus does
not change the unrealistic secondary electron scenario, in which
protons exceed the cluster energy budget.

The work presented here also underlines that observational selection
effects against steep spectrum sources like Abell 521 have likely
missed plenty of similar halo sources, associated with more numerous
but less energetic merging clusters. Current studies on radio halos
are mostly based on observations at 1.4 GHz, where steep spectrum
objects like Abell 521 halos are revealed  much harder than at
lower frequencies (100-300 MHz) 

\acknowledgments

This work is partially supported by INAF under grant PRIN-INAF2007
and by ASI under grant ASI-INAF I/088/06/0.
Basic research in radio astronomy at the Naval Research Laboratory is 
supported by 6.1 base funds.
The National Radio Astronomy Observatory (NRAO) is a facility of the
National Science Foundation operated under cooperative agreement by
Associated Universities, Inc.
The GMRT is run by the National Centre for Radio Astrophysics of the
Tata Institute of Fundamental Research.

\begin{figure}
\epsscale{.90}
\label{uvcov}
%\plotone{figure_0.ps}
\caption[]{
FIGURE AVAILABLE IN GIF FORMAT ONLY

$$- - - -$$

Inner portion of the uv-plane sampled by observations from
which the flux density of the halo has been measured:
{\it top left:} GMRT at 240 MHz; {\it top right:} GMRT at 330 MHz;
{\it bottom left:} GMRT at 610 MHz; {\it bottom right:} VLA at 1400
MHz. Only baselines shorter than 1 k$\lambda$ have been plotted.
}
\end{figure}

\begin{figure}
\epsscale{1.025}
\plotone{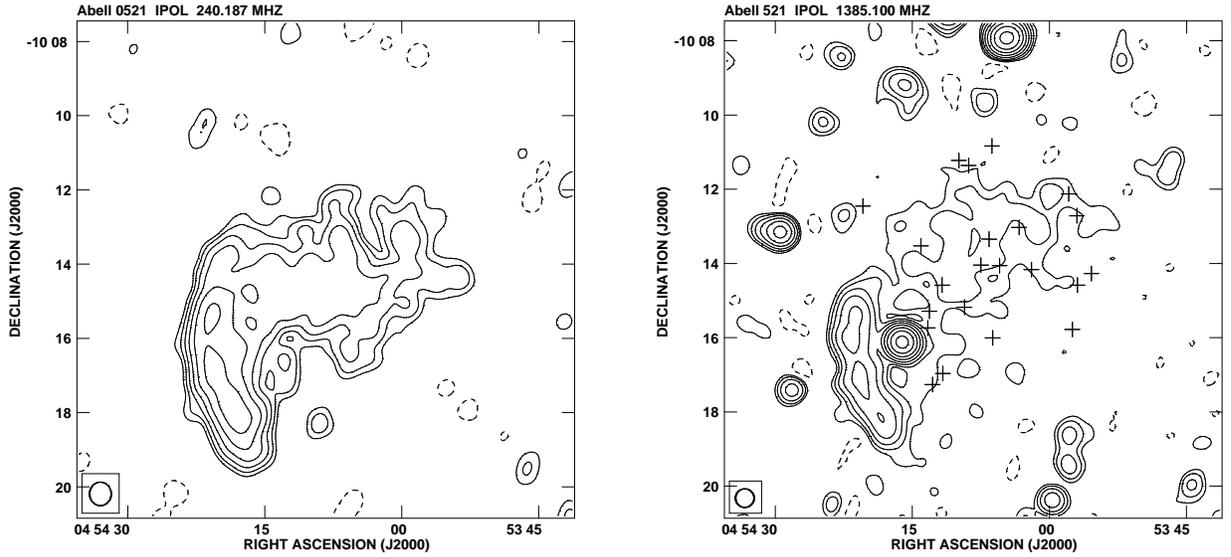}
\caption[]{
Images of the radio halo region: in both cases, contour
levels are -1, 1, 2, 4, 8, 16,  \ldots times the first contour.
{\bf Left}: GMRT 240 MHz contours (data presented in Brunetti et al.
2008). The restoring beam is 38$^{\prime\prime}\times$35$^{\prime\prime}$
in p.a. $0^\circ$ and the first contour is 0.5\,mJy/beam.
{\bf Right}: The new VLA image at 1.4 GHz obtained from a combination
of D and BnC configuration data and restored with a
30$^{\prime\prime}$ circular beam. The r.m.s. noise in the image plane
is about 30$\mu$Jy/beam and the first countour is 85\,$\mu$Jy/beam.
Point sources derived from a full resoltion image (13$^{\prime
\prime}\times$7$^{\prime \prime}$ in p.a. $76^\circ$ , not shown)
have been subtracted out from the region covered by the halo (crosses
mark these locations).
}
\label{fig:profili}
\end{figure}

\begin{figure}
\epsscale{.70}
\plotone{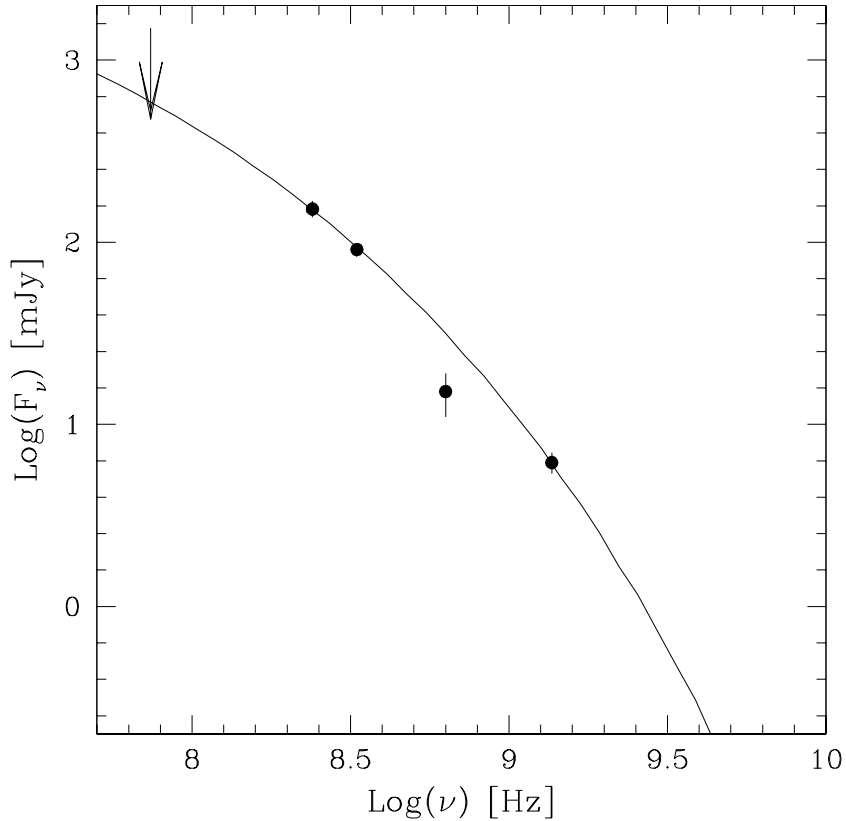}
\caption[]{Synchrotron spectrum of the radio halo in Abell 521.
Measurements at 74, 240, 330, 610 MHz are taken from Brunetti et al.(2008),
the flux density at 1.4 GHz (this paper) is = 6.4$\pm$0.6 mJy.  
The re-acceleration model shown in the figure assumes that
18\% of the thermal energy is in magnetosonic waves.
The synchrotron emission and particle acceleration of low energy
seed relativistic electrons is calculated following Brunetti \& 
Lazarian (2007), assuming physical parameters
of Abell 521 (from Arnaud et al.~2000), a central value of the
magnetic field $B_o=3.5\mu$G and a scaling $B \propto n_{th}$,
$n_{th}$ being the thermal density.
}
\end{figure}

\end{document}